\documentclass[lettersize,journal]{IEEEtran}
\usepackage{stix}
\usepackage{amsmath,amsfonts}
\usepackage{algorithmic}
\usepackage{algorithm}
\usepackage{array}
\usepackage[caption=false,font=normalsize,labelfont=sf,textfont=sf]{subfig}
\usepackage{textcomp}
\usepackage{stfloats}
\usepackage{url}
\usepackage{verbatim}
\usepackage{graphicx}
\usepackage{cite}

\usepackage{commath, color, colortbl, hyperref, soul, wasysym}
\definecolor{tablecolor}{cmyk}{0,0,0,0.12}

\begin{document}

\title{Cross-domain Neural Pitch and \\ Periodicity Estimation}
\author{Max Morrison, Caedon Hsieh, Nathan Pruyne, and Bryan Pardo
\thanks{Max Morrison, Caedon Hsieh, Nathan Pruyne, and Bryan Pardo are with the Department of Computer Science, Northwestern University, Evanston, IL, 60208, USA. (e-mail: morrimax@u.northwestern.edu; pardo@northwestern.edu)}}


\maketitle


\begin{abstract}
Pitch is a foundational aspect of our perception of audio signals. Pitch contours are commonly used to analyze speech and music signals and as input features for many audio tasks, including music transcription, singing voice synthesis, and prosody editing. In this paper, we describe a set of techniques for improving the accuracy of widely-used neural pitch and periodicity estimators to achieve state-of-the-art performance on both speech and music. We also introduce a novel entropy-based method for extracting periodicity and per-frame voiced-unvoiced classifications from statistical inference-based pitch estimators (e.g., neural networks), and show how to train a neural pitch estimator to simultaneously handle both speech and music data (i.e., \textit{cross-domain} estimation) without performance degradation. Our estimator implementations run 11.2x faster than real-time on a Intel i9-9820X 10-core 3.30 GHz CPU---approaching the speed of state-of-the-art DSP-based pitch estimators---or 408x faster than real-time on a NVIDIA GeForce RTX 3090 GPU. We release all of our code and models as Pitch-Estimating Neural Networks (\texttt{penn}), an open-source, pip-installable Python module for training, evaluating, and performing inference with pitch- and periodicity-estimating neural networks. The code for \texttt{penn} is available at \texttt{\href{https://github.com/interactiveaudiolab/penn}{github.com/interactiveaudiolab/penn}}.
\end{abstract}




\section{Introduction}

Vibrating objects produce sound. When those vibrations are periodic (and in the range of human hearing), we perceive a \textit{pitch} related to the fundamental frequency (F0) \cite{yost2001fundamentals}\footnote{Perceived pitch and fundamental frequency are strongly correlated but can occasionally disagree~\cite{hartmann2004signals}. As with prior works, we assume these disagreements can be ignored and refer to both as ``pitch''.}. This perception of pitch can be reduced by adding aperiodic noise, which lowers the \textit{periodicity} of the signal (i.e., the extent to which a segment of audio contains repetition at a regular interval). The human auditory system uses pitch continuity to keep track of individual speakers~\cite{darwin1975dynamic} and parse audio scenes \cite{bregman1994auditory}. Pitch is also central to spoken communication. In languages such as English and Spanish, pitch is used to indicate emphases and phrase boundaries~\cite{nooteboom1997prosody}, and in tone-based languages (e.g., Mandarin Chinese) it indicates lexical content. In music, pitch is fundamental to our understanding of chords and melody.

Given its central importance to speech and music, estimated pitch contours are widely used in music transcription \cite{benetos2018automatic}, music information retrieval~\cite{dannenberg2007comparative, kotsifakos2012survey}, and speech analysis \cite{wennerstrom2001music}. They are also used to train or evaluate systems that perform audio editing and generation tasks, such as speech synthesis~\cite{valin2019lpcnet, ren2020fastspeech, morrison2022chunked, lee2022bigvgan}, music synthesis~\cite{engel2020ddsp}, singing voice synthesis~\cite{liu2022diffsinger}, voice conversion~\cite{qian2020f0}, voice privacy~\cite{o2022voiceblock}, and prosody editing~\cite{morrison2020controllable}. 

In addition to pitch, periodicity contours are also widely used for speech synthesis~\cite{valin2019lpcnet,cho2024articulatory,morrison2024fine} as well as evaluating speech reconstruction accuracy~\cite{morrison2022chunked, lee2022bigvgan}. Periodicity is used for making binary decisions about whether speech is voiced (low periodicity indicates an unvoiced region) or which portion of the pitch contour calculated on input speech is likely to be meaningful (e.g., omitting unpitched regions when evaluating pitch distance).

\begin{figure*}
    \centering
    \includegraphics[width=.9\linewidth]{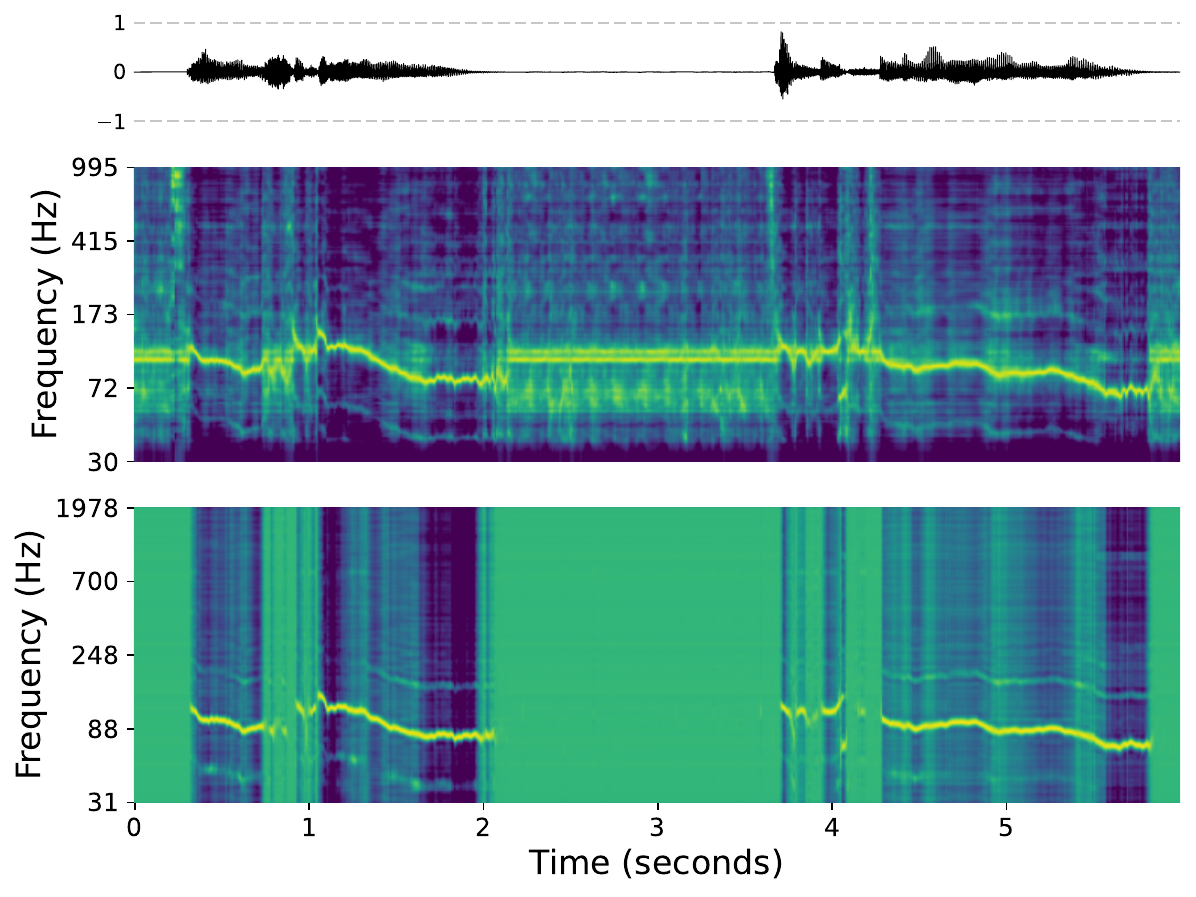}
    \caption{Pitch posteriorgrams produced by our reimplementation of the baseline FCNF0 pitch tracker~\cite{ardaillon2019fully} (\textbf{middle}) and our proposed FCNF0++ (\textbf{bottom}). The input audio (\textbf{top}) is the speech utterance ``I am sitting in a room, different from the one you are in now''. We apply softmax to each time frame to produce normalized distributions and take the natural log for better visualization. Greater brightness indicates higher probability. The y-axis frequency ranges are representative of the pitch bin ranges of the baseline and proposed models. Our proposed methods produce a sharper peak during pitched frames and encourage uniform probability in unpitched regions, making it easy to identify these regions algorithmically (Section~\ref{sec:periodicity}).}
    \label{fig:posterior}
\end{figure*}


Pitch estimation is most commonly conducted using digital signal processing (DSP) methods. Some of the most successful DSP-based pitch estimation methods to date include YIN~\cite{de2002yin}, probabilistic YIN (PYIN)~\cite{mauch2014pyin}, and PEFAC~\cite{gonzalez2014pefac}. DSP-based pitch estimation typically uses a \textit{candidate generating} function (e.g., autocorrelation~\cite{dubnowski1976real}, cumulative mean-normalized difference~\cite{de2002yin}, or normalized cross-correlation~\cite{talkin1995robust}). For a given frame of audio (e.g., a ten millisecond window), a candidate generating function produces a score for each of the (potentially quantized) pitch values within the frequency range of the estimator, indicating the relative likelihood that a pitch is present at that frequency.  This is followed by a heuristic \textit{candidate selection} function that selects the top F0 candidate at each time step, producing a monophonic pitch contour. Typical choices for this candidate selection function are argmax or Viterbi decoding \cite{forney1973viterbi}.

Periodicity is often computed as the per-frame maximum of the candidate generating function, or as the sum of the scores of candidates above some threshold (i.e., the sum of peaks selected via peak-picking)~\cite{mauch2014pyin}.  Existing approaches to periodicity estimation exhibit low voiced/unvoiced accuracy. Some also fail to handle silence, requiring a post-processing step of setting the periodicity of frames with volume below some threshold to zero. 

Pitch estimation using machine learning (ML) outperforms DSP-based methods in accuracy and noise robustness~\cite{han2014neural, kim2018crepe, nakano2019joint, kato2019statistical, airaksinen2019data}. These ML methods use the same candidate selection functions as DSP-based methods (e.g., argmax or Viterbi decoding), but use as the candidate generating function a data-driven model (most often a neural network), instead of DSP-based techniques such as autocorrelation. For the majority of ML-based methods, the candidate generating function produces a valid distribution over the set of possible pitch values (i.e., the posterior distribution inferred by the model). We call this time-varying distribution produced from a sequence of audio frames the \textit{pitch posteriorgram} (Figure~\ref{fig:posterior}). Despite their robustness to noise, existing neural pitch estimators are slow relative to DSP-based methods, and have not shown good generalization across data domains.


We make the following contributions in this paper.

\begin{itemize}
        \item Methods (Section~\ref{sec:pitch}) that improve pitch and periodicity estimation of several top recent neural pitch estimators.
    \item A neural pitch estimator, built using these methods, that is capable of fast and accurate pitch and periodicity estimation on music and speech data.
    \item A novel entropy-based method for extracting per-frame signal periodicity that improves framewise voiced/unvoiced classification of speech (Section~\ref{sec:periodicity}).
\end{itemize}
A thorough evaluation of all proposed methods and models demonstrates that our proposed approach produces state-of-the-art pitch and periodicity estimation and that neural pitch estimation can approach the CPU inference speed of DSP-based methods. We release our code as Pitch-Estimating Neural Networks (\texttt{penn}), a pip-installable Python library containing all of our training, evaluation, and inference code and as well as pretrained models for pitch and periodicity estimation. The code for \texttt{penn} is available at \texttt{\href{https://github.com/interactiveaudiolab/penn}{github.com/interactiveaudiolab/penn}}.


\section{Baseline architectures}
\label{sec:baselines}
This work builds on the standard neural pitch tracking paradigm, where a statistical model (the candidate generating function) takes a frame of audio as input and outputs a posterior distribution over the set of possible pitches (Figure~\ref{fig:posterior}) and then a candidate selection algorithm selects the best pitch estimate at each time step, using the sequence of probability distributions as input. Perhaps the most most widely-used pitch tracker of this kind is CREPE~\cite{kim2018crepe}. We use CREPE as well as two pitch trackers that build upon CREPE as baselines to demonstrate the efficacy of our methods. We now describe these three baseline systems.


\subsection{CREPE}
\label{sec:crepe}

Convolutional REpresentation for Pitch Estimation (CREPE) is a neural pitch estimator composed of six convolutional blocks followed by a linear layer. Each convolutional block consists of a one-dimensional convolution, ReLU activation, batch normalization~\cite{ioffe2015batch}, and dropout~\cite{srivastava2014dropout}. CREPE takes as input 1024 samples of a 16 kHz audio waveform that has been preprocessed to have a mean of zero and standard deviation of one. The network produces logits that yield independent Bernoulli posterior probabilities over each pitch bin after sigmoid normalization.
\begin{equation}
    p(y = c_i | x), \; i = 0, 1, \dots, P - 1,
\end{equation}

\noindent
where $x \in \mathbb{R}^{1024}$ are the input audio samples, $y$ is the pitch of the input audio, $c_i$ is the center frequency of quantized pitch bin $i$, and $P$ is the number of pitch bins. CREPE uses $P = 360$ pitch bins, spaced every 20 cents between 31 and 2006 Hz.
\begin{equation}
    c_i = 31 \times 2^{\frac{20 i}{1200}}
\end{equation}

\noindent
CREPE is trained in a supervised manner. The probability for each pitch bin is predicted independently due to the choice of loss function. Rather than transforming the network logits to a categorical distribution over pitch bins via a softmax activation and evaluating using categorical cross entropy (CCE) loss, CREPE uses a sigmoid activation and binary cross-entropy (BCE) loss on each pitch bin. This choice means that the network can simultaneously place high confidence at multiple pitch bins. This could be useful for polyphonic pitch estimation; however, our work focuses on monophonic pitch estimation. CREPE applies a Gaussian blur to the one-hot-encoded training targets (with a standard deviation of 25 cents) followed by peak normalization. This encourages the network to increase the variance of its prediction by assigning probability mass to adjacent pitch bins.

The pitch candidate selection algorithm used in CREPE produces pitch values (in Hz) from the quantized output space by removing all probability density outside of a window of nine pitch bins (180 cents) centered on the argmax pitch bin and then taking the expected value. We call this \textit{local expected value decoding}, as it improves resolution by assuming the expected value of the normalized distribution mass near the argmax prediction coincides with the peak.


\subsection{DeepF0}
\label{sec:deepf0}

DeepF0~\cite{singh2021deepf0} makes only architectural changes to CREPE so that the data preprocessing, quantization, batch size, loss function, early stopping, and sampling rates remain the same. The authors of DeepF0 propose dilated convolutions in order to increase the receptive field, as well as residual connections to facilitate faster training. The architecture of DeepF0 consists of a one-dimensional causal convolution, four residual blocks, a one-dimensional average pooling, and a linear layer over the pitch bin categories. Each residual block consists of a dilated, causal one-dimensional convolution, ReLU activation, and a weight-normalized~\cite{salimans2016weight} one-dimensional convolution with a kernel size of one. The input to each block is added to the output and passed through a ReLU activation. All other kernel sizes are 64 and all hidden channel sizes are 128. Dilation rates in the four residual blocks increase exponentially: 1, 2, 4, and 8.


\subsection{FCNF0}
\label{sec:fcnf0}

Fully-Convolutional Network for Pitch Estimation (FCNF0)~\cite{ardaillon2019fully} makes several modifications to CREPE. FCNF0 omits zero-padding from convolutional layers, as a significant amount of inference time in CREPE is spent performing computations on zero-padding. The authors replace the output linear layer with a convolutional layer to improve speed and further improve CPU throughput with a fully-convolutional inference mode. FCNF0 omits dropout to increase the network capacity and uses an 8 kHz sampling rate for the input audio instead of 16 kHz. Otherwise, the six convolutional blocks of FCNF0 are identical to CREPE, but with a different number of convolution channels, and with pooling strides and kernel sizes hand-tuned to produce output of a desired length provided the reduction in the time dimension due to omitting zero-padding. The output categories of FCNF0 are 486 pitch bins with a minimum pitch of 30 Hz and a bin width of 12.5 cents, resulting in a frequency range of 30-1000 Hz.



\subsection{Training Approach}
CREPE, FCNF0, and DeepF0 are all trained with a batch size of 32 until 16,000 steps have passed in which the validation accuracy has not improved (i.e., ``early stopping''). We assume the validation batch size is the same as training and performed an extensive search over the number of validation batches and whether the validation data is shuffled after each epoch to replicate the performance of CREPE on the MDB-stem-synth dataset. For our reimplementations, we use 64 validation batches without shuffling after each epoch, so that roughly 10\% of training time is devoted to validation. We utilize this early stopping method for our baseline implementations of CREPE, FCNF0, and DeepF0.


\section{Improving pitch and periodicity estimation}
\label{sec:pitch}

In this section, we describe a series of improvements applied to our three baseline neural pitch estimators (CREPE, DeepF0, and FCNF0) to improve pitch and periodicity estimation. We call resulting pitch trackers CREPE++, DeepF0++, and FCNF0++. For details on the incremental and cumulative improvements resulting from proposed changes to the baseline systems, see Tables \ref{tab:pitch} and \ref{tab:ablate}.

\subsection{Proposed best practices}
\label{sec:new-methods}

\noindent
\textbf{Increase the frequency resolution $\vert$} CREPE and DeepF0 predict bins with a quantization width (i.e., the spacing between the centers of two adjacent bins) of 20 cents. FCNF0 predicts bins with a width of 12.5 cents. Assuming a uniform distribution of noiseless, continuous ground truth pitch values over the range of a given pitch bin, the expected value of the error of the quantization of CREPE or DeepF0 is five cents. To see this, note that the minimum error is zero and the maximum error is one half-bin. The expected value of a uniform distribution between real values $a$ and $b$, where $b > a$, is $(b - a) / 2$ (i.e., one quarter-bin). We reduce the bin width to five cents (i.e., 1440 pitch bins), which reduces the expected quantization error to 1.25 cents.

\noindent
\textbf{Train on frames without a pitch $\vert$} Prior methods were trained using only frames in which the ground truth annotations indicate that a pitch is present. However, a proper periodicity estimator must behave sensibly for audio without periodic structure (i.e., providing a low score). We include frames without pitch labels during training, setting their ground truth pitch bin to a random bin. This promotes a high-entropy uniform distribution in aperiodic regions, which we make use of for periodicity estimation in Section~\ref{sec:periodicity}.

\noindent
\textbf{Do not stop early $\vert$} Our baseline models are undertrained as a result of early stopping (Section~\ref{sec:crepe}). Our baseline models all stop when the validation accuracy has not improved for 16k steps, where evaluation occurs every 500 steps. This leads to significantly fewer training steps than training to convergence. Running training for an indefinite period, we find that the model converges closer to (or slightly before) 250k steps.

\noindent
\textbf{Do not normalize the input $\vert$} All baseline models normalize the input audio to have a mean of zero and a standard deviation of one. While this improves performance early in training, removing early stopping demonstrates that normalization slightly harms pitch and periodicity accuracy and marginally increases the amount of compute required. We omit this normalization.

\noindent
\textbf{Replace binary cross entropy loss $\vert$} Binary cross entropy independently predicts Bernoulli distributions for each pitch bin at each time step. Categorical cross entropy (CCE) predicts a single categorical distribution over all pitch bins at each time step. CCE improves performance metrics when early stopping is omitted. It is also a more sensible choice when used with our entropy-based periodicity measure (Section ~\ref{sec:periodicity}), which assumes the input is a categorical distribution. We apply the same Gaussian blur used in training CREPE (Section \ref{sec:crepe}) when training all models, with a standard deviation of 25 cents applied to the training targets. This improves performance for both BCE and CCE losses.

\noindent
\textbf{Increase the batch size $\vert$} Prior methods use normalization methods such as batch and weight normalization. Batch normalization induces instabilities when the batch size is small~\cite{ba2016layer}. Increasing the batch size can help to remove these instabilities. We verify this by increasing the batch size from 32 to 128, where a batch size of 128 is the largest multiple of two within the memory capacity of our GPU for all models.

\noindent
\textbf{Use layer normalization $\vert$} Increasing the batch size may help, but batch normalization is still non-robust to outliers. We use layer normalization~\cite{ba2016layer} to solve this issue and improve performance relative to batch or weight normalization. In other words, rather than normalizing the network weights or normalizing over the batch and length dimensions, we normalize over the channel and length dimensions.


\subsection{Proposed best practices from prior works}
\label{sec:prior-methods}

We now describe methods that were used in at least one of the original baseline systems and that we apply in all of our proposed models (CREPE++, DeepF0++, and FCNF0++).

\noindent
\textbf{Remove dropout $\vert$} CREPE~\cite{kim2018crepe} uses dropout~\cite{srivastava2014dropout} to prevent overfitting. However, the network does not overfit when dropout is removed---in fact, performance and training speed improve. This indicates that the model was able to utilize additional capacity that was constrained by dropout. This was previously noted by the authors of DeepF0~\cite{singh2021deepf0} and FCNF0~\cite{ardaillon2019fully}.

\noindent
\textbf{Use local expected value decoding $\vert$} The local expected value decoding introduced by CREPE slightly improves pitch accuracy relative to argmax decoding. Local expected value decoding also solves an issue called \textit{banding}, which can occur during requantization of an already quantized signal. For example, when using argmax decoding, the resulting pitch is still quantized. Downstream applications that perform requantization of the pitch contour at a different resolution as a preprocessing step can decrease the effective resolution if the quantization bins are misaligned. \textit{Banding} refers to this issue of compounding quantization error. While this can be solved via \textit{dithering}, which trades off quantization error for noise~\cite{vanderkooy1987dither}, we find that local expected value decoding is a fast and more accurate alternative.

\noindent
\textbf{Downsample $\vert$} The authors of FCNF0~\cite{ardaillon2019fully} propose using an 8 kHz sampling rate for all training and inference instead of the 16 kHz sampling rate used in CREPE~\cite{kim2018crepe} and DeepF0~\cite{singh2021deepf0}. We notice significant improvements in pitch and periodicity estimation accuracy at 8 kHz relative to 16 kHz due to the larger context captured by the constant window size during both training and inference. These improvements outweigh the detrimental effects of losing harmonic information. Therefore, we use an 8 kHz sample rate in our proposed systems.


\section{Periodicity estimation with statistical pitch estimators}
\label{sec:periodicity}

Periodicity estimation provides a real-valued score for the extent to which each frame of audio contains a pitch. Such periodicity contours are commonly used for the input~\cite{morrison2024fine,cho2024articulatory} and evaluation~\cite{morrison2022chunked,lee2022bigvgan} of speech synthesis systems. Periodicity estimation with statistical pitch estimators has been treated inconsistently in recent literature on neural pitch estimation. Some studies omit the evaluation of periodicity or voicing~\cite{kim2018crepe, kato2019statistical}. Others demonstrate binary voicing classification that---at best---slightly outperforms DSP-based baselines~\cite{tran2020robust, gfeller2020spice}. In this work, we apply two methods for estimating a periodicity value $h \in \left[ 0, 1 \right]$ for input audio $x$ from categorical posterior distributions of neural pitch estimators, where a periodicity near zero indicates aperiodic noise and a periodicity near one indicates a noise-free, pitched signal.

\noindent
\textbf{Method one (\textit{max})} is a simple, domain-agnostic baseline approach that takes the maximum posterior probability over the pitch bins in each time frame.
\begin{equation}
    \hat{h}^{(\text{max})} = \max_i p(y = c_i | x)
\end{equation}

\noindent
\textbf{Method two (\textit{entropy})} is our novel method that derives periodicity from the entropy of the categorical posterior distribution. The periodicity is scaled to the range $\left[ 0, 1 \right]$ by dividing by $\ln P$---the maximum entropy of a categorical distribution with $P$ categories---and subtracted from 1 (i.e., low entropy indicates high periodicity).
\begin{equation}
    \hat{h}^{(\text{entropy})} = 1 - \frac{1}{\ln P} \sum_{i=0}^{P - 1} p(y = c_i | x) \ln p(y = c_i | x)
\end{equation}

\noindent
Unlike the direct prediction method of Gfeller et al.~\cite{gfeller2020spice}, our method requires no additional loss functions or modifications to the neural network architecture. This reduces complexity and improves training and inference speed. Unlike baseline method one (\textit{max}), our entropy-based method can elegantly handle polyphony. Consider a peak-normalized sine wave with a frequency within the range of the pitch estimator. This signal should produce a periodicity of one. Now, add another sine wave at a different frequency also within the range of the estimator, so that the posterior distribution produced by the model has two peaks. The resulting signal should still have a periodicity of one. Method one (\textit{max}) produces a periodicity of 0.5, while our proposed entropy-based method produces a periodicity of $1 -\frac{1}{\ln{P}} (2 \times 0.5 \times \ln 0.5)$. For our proposed $P = 1440$, this produces a periodicity of 0.905. In this example, as $P$ approaches infinity, the periodicity approaches one, as desired.


\subsection{Binary voicing decisions}
\label{sec:binary}

Once we have produced a periodicity estimate, we can perform thresholding to produce per-frame classifications of whether a pitch is present. This is referred to as the voiced/unvoiced decision in the context of speech. Specifically, we aim to make a sequence of binary voiced/unvoiced decisions $\hat{v}_0, \dots, \hat{v}_{n-1}$ from predicted pitch contour $\hat{y}_0, \dots, \hat{y}_{n-1}$ and corresponding periodicities $\hat{h}_0, \dots, \hat{h}_{n-1}$. We use the following decision rule, where $\alpha \in \left[ 0, 1 \right]$ is a voicing threshold hyperparameter.
\begin{equation}
    \hat{v}_t = \begin{cases}
      1, & \text{if}\ \hat{h}_t > \alpha \\
      0, & \text{otherwise}
    \end{cases}
\end{equation}


\section{Evaluation}
\label{sec:evaluation}

We design our evaluation to test two hypotheses: (1) our proposed methods improve the pitch and periodicity accuracy of the widely-used neural pitch estimators we build upon and (2) neural pitch estimators can generalize across data domains of speech and music. We further show that our models approach the CPU inference speed of state-of-the-art DSP-based pitch estimators. Finally, we compare our pitch estimation results to prior results from another recent approach (HarmoF0)~\cite{wei2022harmof0} that has slightly improved pitch-tracking performance relative to our baseline methods, but requires a full audio file for processing.

\subsection{Data}
\label{sec:data}

Ground truth pitch annotations for natural signals necessitate complex methods that can induce noise and require humans to make manual corrections. Nonetheless, existing datasets have proven sufficient for training and evaluating pitch estimators that outperform DSP-based methods. We use one music and one speech dataset, each with ground-truth pitch and voicing annotations. For each dataset, we perform a random 70-15-15 data split of files into training, validation, and test partitions.

For music, we use MDB-stem-synth~\cite{salamon2017analysis}, a music dataset used in multiple recent pitch estimation papers. MDB-stem-synth consists of 230 solo stems from MedleyDB~\cite{bittner2014medleydb} resynthesized from ground truth pitch for a total of 15.6 hours of music data. 

PTDB~\cite{pirker2011pitch} is the dataset most commonly used in recent work on pitch estimation for speech. For this reason, we use PTDB as a representation of performance on speech data. PTDB consists of 4718 English speech and corresponding laryngograph recordings (contact microphones placed on the neck) with human-corrected pitch annotations and boolean voiced/unvoiced annotations for a total of 9.6 hours of speech data.

\subsection{Evaluating periodicity and voicing decisions}
\label{sec:periodicity-eval}

We evaluate periodicity via classification F1 score of the binary voiced/unvoiced decision (Section~\ref{sec:binary}). We extract voicing decisions from estimated periodicity using both the \textit{max} and \textit{entropy} periodicity methods (Section~\ref{sec:periodicity}). We report the F1 score of the binary voicing classification for each neural pitch estimation method that we evaluate.

For each voicing threshold hyperparameter search, we use the validation data partitions of both datasets to perform two grid searches using values $\alpha = 0.0, 0.1, \dots, 0.9$ and $\alpha = 2^{-i}, \; i = 1, 2, \dots, 9$, producing candidate threshold value $\alpha^*$. We then assume the hyperparameter landscape is convex at $\alpha^*$ and perform a sequential grid search, reducing the step size from 0.05 by a factor of two at each step for eight steps. This can be thought of as fine-tuning $\alpha^*$ via gradient ascent on the F1 score with an exponentially decreasing step size, where first-differences are used to measure the gradient direction. This fine-grained hyperparameter search is necessary for all models without our proposed unvoiced training strategy. Without this training strategy, models exhibit a narrow peak for the optimal hyperparameter value (Figure~\ref{fig:threshold}).


\begin{table*}[ht]
\centering
\begin{tabular}{l|ccc|cc|cr}
& \multicolumn{3}{c|}{\textbf{Pitch}} & \multicolumn{2}{c|}{\textbf{Periodicity}} & \multicolumn{2}{c}{\textbf{Time}} \\
\textbf{Model} & \textbf{$\boldsymbol{\Delta\cent}\downarrow$} & \textbf{RPA$\uparrow$} & \textbf{RCA$\uparrow$} & \textbf{F1 (Entropy)$\uparrow$} & \textbf{F1 (Max)$\uparrow$} & \textbf{RTF (GPU)$\downarrow$} & \textbf{RTF (CPU)$\downarrow$}\\
\hline
CREPE~\cite{kim2018crepe} & 21.07 & .9748 & .9780 & .9626 & .9509 & .0093 & .3574 \\
\rowcolor{tablecolor}
CREPE++ & 15.15 & .9783 & .9817 & .9801 & .9799 & .0102 & .3632 \\
DeepF0~\cite{singh2021deepf0} & 20.12 & .9778 & .9811 & .9509 & .9239 & .0178 & 1.1581 \\
\rowcolor{tablecolor}
DeepF0++ & \underline{12.66} & \textbf{.9828} & \textbf{.9856} & \underline{.9815} & \textbf{.9814} & .0194 & 1.2078 \\
FCNF0~\cite{ardaillon2019fully} & 18.01 & .9719 & .9748 & .9602 & .9361 & \textbf{.0019} & .0753 \\
\rowcolor{tablecolor}
\textbf{FCNF0++} & 12.72 & \underline{.9825} & .9852 & \textbf{.9816} & \underline{.9813} & \underline{.0024} & .0861 \\
\quad \textbf{+ Local expected value} & \textbf{12.45} & \underline{.9825} & \underline{.9853} & \textbf{.9816} & \underline{.9813} & \underline{.0024} & .0892 \\
\hline
\rowcolor{tablecolor}
\texttt{torchcrepe}~\cite{morrison2022torchcrepe} & 59.40 & .9103 & .9183 & .9293 & .9305 & .0199 & .6435 \\
PYIN~\cite{mauch2014pyin} & 110.5 & .8477 & .8643 & \multicolumn{2}{c|}{$.9199^\dagger$} & - & \underline{.0639} \\
\rowcolor{tablecolor}
DIO~\cite{morise2009fast} + Stonemask~\cite{morise2016world} & 80.10 & .6961 & .7043 & - & - & - & \textbf{.0177} \\
\end{tabular}
\caption{Pitch and periodicity error and speed of our baselines, our proposed models, and common open-source models on both PTDB and MDB-stem-synth datasets. Pitch error in cents ($\Delta\cent$), raw pitch accuracy (RPA), raw chroma accuracy (RCA), voiced/unvoiced F1, and real-time factor (RTF) metrics are described in Sections~\ref{sec:periodicity-eval}-~\ref{sec:speed-eval}. We consider FCNF0++ with local expected value decoding to be most useful model for most downstream applications, with competitive accuracy and speed. \\ $^\dagger$PYIN uses the sum of peak-picked densities for periodicity decoding. \\$\uparrow$ indicates that higher is better and $\downarrow$ indicates that lower is better.}
\label{tab:pitch}
\end{table*}


\subsection{Evaluating pitch}
\label{sec:pitch-eval}

We measure the pitch error of voiced frames using the average error in cents ($\Delta \cent$), raw pitch accuracy (RPA), and raw chroma accuracy (RCA). Cents is a perceptually useful measure of pitch error due to its correspondence with musical intervals: 100 cents is one semitone. Let $\cent(y, \hat{y}) = \abs{1200 \log_2 (y / \hat{y})}$ be the absolute difference in cents between ground-truth pitch $y$ and estimated pitch $\hat{y}$, both measured in Hz. Let $\epsilon$ be a pitch threshold in cents. The RPA and RCA are calculated as follows.
\begin{equation}
    RPA_{\epsilon} = \frac{1}{n} \sum_{t=0}^{n - 1} \Big( \cent \left( y_t, \hat{y}_t \right) < \epsilon \Big)
\end{equation}

\begin{equation}
    RCA_{\epsilon} = \frac{1}{n} \sum_{t=0}^{n - 1} \Big( \Big( \cent \left( y_t, \hat{y}_t \right) \bmod 1200 \Big) < \epsilon \Big)
\end{equation}

\noindent
In other words, RPA is the fraction of frames with pitch error greater than $\epsilon$ cents. RCA does not penalize octave shifts. Therefore, the gap between RPA and RCA is an indicator of half- and double-frequency errors. We use a pitch threshold of $\epsilon = 50$ cents.


\subsection{Evaluating speed}
\label{sec:speed-eval}

Fast pitch estimation is imperative for processing large datasets as well as for real-time applications. We evaluate the inference speed of each pitch estimator using both CPU and GPU compute, reporting the real-time factor (RTF), or the average number of seconds it takes to perform pitch and periodicity estimation on one second of audio. We use a hopsize of ten milliseconds. We use one Intel i9-9820X 3.30 GHz 10-core CPU for CPU inference and one NVIDIA GeForce RTX 3090 for GPU inference. We use a batch size of up to 2048 frames during GPU inference, requiring multiple forward passes for audio files greater than 2048 frames (20.48 seconds of audio). We do not limit the batch size for CPU inference. RTF values are shown in Table \ref{tab:pitch}. We do not provide GPU speeds for our baseline DSP-based pitch estimators that do not provide GPU implementations. All reported speeds include loading of audio, copying between devices, and saving results. We do not provide a benchmark of real-time streaming-based pitch estimation of our models, but note that this is possible with our methods and the latency of streaming-based inference depends on the batch size.


\subsection{Experiments}

\noindent
\textbf{Improved pitch and periodicity estimation} We apply our proposed methods (Section~\ref{sec:pitch}) to each of our three baseline models (Section~\ref{sec:baselines}), producing CREPE++, DeepF0++, and FCNF0++. We perform training and inference on both MDB-stem-synth and PTDB, and use both \textit{max} and our proposed \textit{entropy} methods for periodicity decoding (Section~\ref{sec:periodicity}). Unless otherwise specified, we use argmax for pitch decoding for all models instead of local expected value decoding (Section~\ref{sec:prior-methods}) in order to avoid having to perform a hyperparameter search on the window size for every model. 

We find FCNF0++ to be our most broadly-useful model, since its pitch accuracy is roughly equal to the best model for pitch (DeepF0++), its periodicity estimation is equal or better, and it runs an order of magnitude faster than DeepF0++. We separately evaluate the impact of local expected value decoding on FCNF0++ with a window size of 19 (i.e., 95 cents) determined via hyperparameter search over odd window sizes. We compare to our baseline models as well as three common open-source pitch estimators: PYIN~\cite{mauch2014pyin}, an accurate DSP-based method; DIO~\cite{morise2009fast}, a fast DSP-based method written primarily in C++; and the open-source implementation of CREPE trained on six music datasets. While PYIN is often used in combination with Viterbi pitch decoding, we instead use argmax decoding for fair comparison with our other methods. We use the \texttt{torchcrepe}~\cite{morrison2022torchcrepe} wrapper of CREPE with argmax decoding. This produces the same results as the original Keras implementation~\cite{kim2022crepe} up to floating point error in the input normalization. We perform estimation using PYIN and DIO using a 16 kHz sampling rate, which we find outperforms an 8 kHz sampling rate for these estimators. The open-source implementation of DIO does not provide a method for estimating periodicity. We perform ablations of each of our proposed methods on our FCNF0++ model to assess their relative merit.

\begin{table*}[ht]
\centering
\begin{tabular}{l|ccc|c}
& \multicolumn{3}{c|}{\textbf{Pitch}} & \textbf{Periodicity} \\
\textbf{Model} & \textbf{$\boldsymbol{\Delta\cent}\downarrow$} & \textbf{RPA$\uparrow$} & \textbf{RCA$\uparrow$} & \textbf{F1 (Entropy)$\uparrow$}\\
\hline
FCNF0++ & \underline{12.72} & \textbf{.9825} & \textbf{.9852} & \textbf{.9816} \\
\rowcolor{tablecolor}
\quad Coarse quantization & 17.23 & .9771 & .9798 & .9814 \\
\quad Voiced only & \textbf{12.62} & .9815 & \underline{.9845} & .8496 \\
\rowcolor{tablecolor}
\quad Early stopping & 13.32 & .9800 & .9830 & .9804 \\
\quad Input normalization & 13.14 & \underline{.9822} & \textbf{.9852} & .9811 \\
\rowcolor{tablecolor}
\quad Binary cross-entropy loss & 13.64 & .9817 & \underline{.9845} & \underline{.9815} \\
\quad Smaller batch size & 13.63 & .9792 & .9822 & .9800 \\
\rowcolor{tablecolor}
\quad Batch normalization & 17.44 & .9789 & .9824 & .9734 \\
\rowcolor{tablecolor}

\end{tabular}
\caption{Ablations of our proposed methods described in Section~\ref{sec:pitch}. For example, ``Early stopping`` is FCNC0++ trained with early stopping and ``Voiced only`` is FCNF0++ trained only on voiced frames. Rows are not cumulative: each row independently evaluates exactly one of our suggested improvements (see Section \ref{sec:pitch}) relative to our proposed FCNF0++ model. All models are trained and evaluated on both PTDB and MDB-stem-synth datasets. Pitch error in cents ($\Delta\cent$), raw pitch accuracy (RPA), raw chroma accuracy (RCA), and voiced/unvoiced F1 metrics are described in Sections~\ref{sec:periodicity-eval}-~\ref{sec:pitch-eval}. \\$\uparrow$ indicates that higher is better and $\downarrow$ indicates that lower is better.}
\label{tab:ablate}
\end{table*}


\noindent
\textbf{Cross-domain generalization} We train FCNF0++ on MDB-stem-synth and PTDB individually as well as jointly. We do this to examine how training exclusively on one dataset impacts generalization on the other, when compared to training on both. We investigate any failures to generalize from one dataset to another by comparing the training data distributions over pitch bins to produce insights of desirable properties of the training data of a neural pitch and periodicity estimator.


\section{Results}
\label{sec:results}

We discuss our experimental results in context with our evaluation hypotheses.

\noindent
\textbf{Our proposed methods improve pitch and periodicity estimation}
Results for pitch and periodicity estimation accuracy and speed for our baseline models, our proposed models, and additional baseline neural- and DSP-based pitch and periodicity estimators are found in Table~\ref{tab:pitch}. We see that our proposed methods reduce the pitch estimation error in cents of all baseline models by at least 28\% and the entropy-based periodicity estimation error by at least 46\%.
Other than using automatic mixed precision (AMP), we do not perform any compute optimizations such as operator fusion via graph compilation---as in Pytorch 2.0, asynchronous loading and saving of data, or Open Neural Network Exchange (ONNX)~\cite{bai2019onnx} model exporting. FCNF0++ exhibits highly competitive pitch and periodicity estimation, CPU inference speeds approaching DSP-based methods, and GPU inference speeds that can process large audio datasets in minutes.

\begin{figure}[t]
\centering
\includegraphics[width=\linewidth]{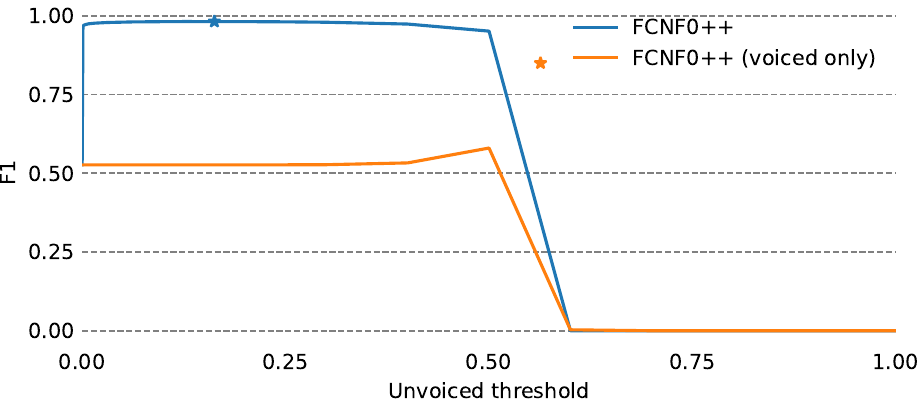}
\caption{Hyperparameter landscape of the voiced/unvoiced threshold on the entropy-based periodicity estimate produced by FCNF0++ with (blue) and without (orange) our proposed unvoiced training strategy (Section~\ref{sec:new-methods}) on PTDB and MDB-stem-synth. Stars indicate optimal F1 values found via a fine-grained binary search (Section~\ref{sec:periodicity-eval}). Notice that---without unvoiced training strategy---the peak is narrow and can only be found via the fine-grained search (i.e., the orange star is not on the orange line). Our unvoiced training strategy of selecting a random bin (Section~\ref{sec:new-methods}) improves the optimal F1 score of the model and produces state-of-the-art voiced/unvoiced classification F1 scores across a large region of the hyperparameter space.}
\label{fig:threshold}
\end{figure}


\begin{figure*}[t]
\centering
\includegraphics[width=\textwidth]{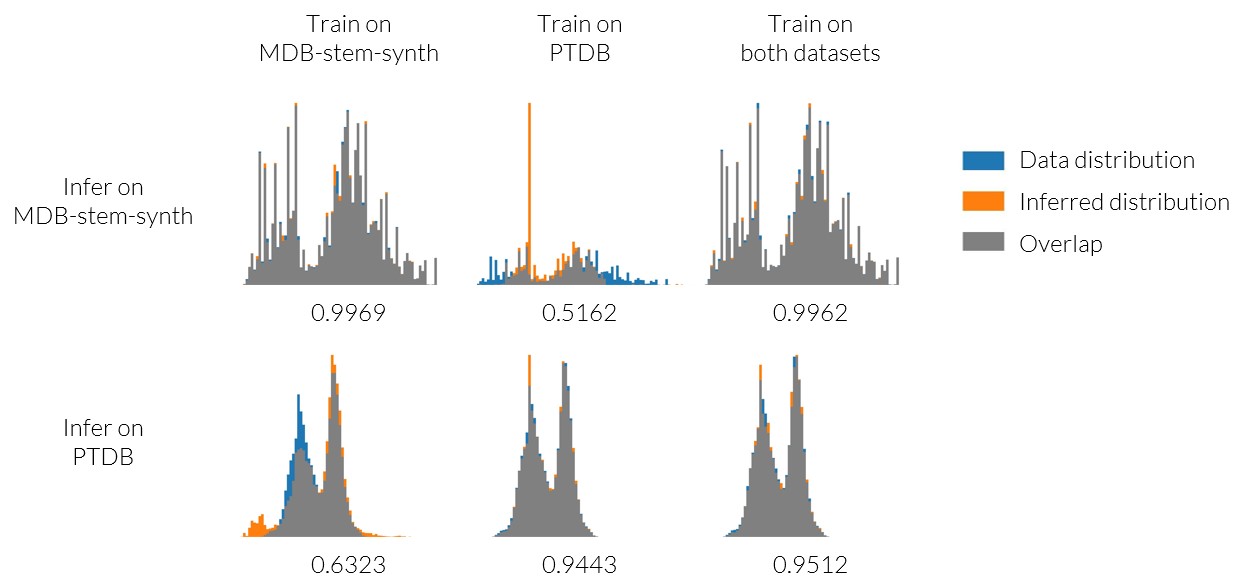}
\caption{\textbf{Cross-domain generalization across music and speech} $|$ Test data distributions (blue), FCNF0++ model inference distributions (orange), and their overlap (gray) for the pitch bins of the test partitions of MDB-stem-synth (top) and PTDB (bottom). In other words, blue indicates where the model should have placed probability and did not, orange indicates where the model placed probability where it should not, and gray indicates the model is correct. The horizontal axis of each set of distributions is pitch bins from low (left) to high (right). \textbf{Raw pitch accuracies (RPAs) with a 50 cent threshold (see Section~\ref{sec:pitch-eval}) are provided below each set of distributions.}}
\label{fig:density}
\end{figure*}


Table~\ref{tab:ablate} shows the pitch and periodicity accuracy of ablations of our proposed improvements relative to our FCNF0++ model (see Section \ref{sec:pitch}). All of our proposed methods improve both pitch and periodicity accuracy. The pitch accuracy of using only voiced frames for training is comparable; however, our proposed method for training on unpitched frames by selecting a random ground truth pitch bin substantially improves periodicity accuracy. The significance of this effect is visible in Figure~\ref{fig:threshold}. Without our unvoiced training strategy, the optimal F1 score decreases and the landscape of unvoiced thresholds with a competitive voiced/unvoiced classification F1 score narrows. According to Tables~\ref{tab:pitch} and~\ref{tab:ablate}, our most notable improvements are our unvoiced training strategy and entropy-based periodicity, as well as using a smaller quantization width and layer normalization.

We briefly compare our results to the HarmoF0 model proposed by Wei et al.~\cite{wei2022harmof0}. HarmoF0 is non-causal, requiring access to the full audio file. This prohibits use for low-latency or streaming-based inference. HarmoF0 is individually trained and evaluated on MDB-stem-synth~\cite{salamon2017analysis} and then trained and evaluated on PTDB~\cite{pirker2011pitch} (i.e., it is not cross-domain and the same model was never trained on both domains). They report RPA values of 98.46\% for music (MDB-stem-synth) and 94.59\% for speech (PTDB). To the best of our knowledge, these are the previous state-of-the-art RPA values on these datasets. These values can be compared (up to variations in the test partition) to the individual-domain RPA values reported in Figure~\ref{fig:density} (i.e., 99.69\% music and 94.43\% speech). Adding cross-domain training changes our RPAs to 99.62\% music  and 95.12\% speech, making our proposed system state-of-the-art while permitting low-latency and streaming-based inference.

\noindent
\textbf{Neural pitch estimators generalize across data domains}
Figure~\ref{fig:density} demonstrates that neural pitch estimators can perform accurate cross-domain pitch and periodicity estimation when the distribution of training data sufficiently covers the frequency range of the evaluation data. Assuming the training partition of each dataset covers the corresponding test partition, training on both datasets guarantees this coverage. To see this, observe that models trained on MDB-stem-synth perform poorly on the male speakers of PTDB (bottom left). This may be due to a gap in the data distribution in MDB-stem-synth around those frequencies. Models trained on PTDB perform poorly on MDB-stem-synth (top middle) due to a lack of coverage of low and high frequencies. Models trained on both datasets (right) perform well on both datasets. For brevity, we display results only for our best estimator, FCNF0++. CREPE++ and DeepF0++ exhibit similar cross-domain generalization. One can conclude that it is desirable for the training data distribution of a neural pitch estimator to at least cover the distribution used during inference. If the inference distribution is unknown---a common setting for pitch and periodicity estimators---it is advantageous to train using a high-entropy (e.g., uniform) distribution. Investigating this setting of universal neural pitch estimation using a high-entropy training distribution that generalizes across datasets without retraining is a natural next step to our current work.


\section{Conclusion}
\label{sec:conclusion}

Pitch and periodicity contours are necessary representations for many audio tasks. Our work significantly advances the state-of-the-art of pitch and periodicity estimation of speech and music. We demonstrate a set of methods that improve pitch and periodicity estimation performance, including a training method for unpitched audio frames and a more fine-grained pitch quantization. We introduce an entropy-based periodicity decoding method that improves periodicity estimation. We perform thorough evaluation of our proposed methods and demonstrate neural pitch and periodicity estimation speeds that approach the CPU estimation speeds of state-of-the-art DSP-based methods. All experiments described in this paper are reproducible with configuration files found in Pitch-Estimating Neural Networks (\texttt{penn}), our open-source Python library.

Future work in pitch and periodicity estimation may seek to outperform DSP-based methods in CPU inference speed, utilize higher-entropy training data distributions, demonstrate universal generalization on unseen datasets, and perform polyphonic pitch and periodicity estimation via either source separation or peak picking of the pitch posteriorgram. It is also worth investigating the relationship between periodicity and signal energy to determine whether correlation between these representations is detrimental for downstream applications. If so, it would be useful to address this correlation and produce a periodicity estimate that is independent of signal energy. Lastly, advances in neural pitch estimation quality may permit effective differentiable pitch supervision by penalizing a generative model for producing audio with the incorrect pitch or periodicity using the gradients of a pitch estimator.


\bibliographystyle{IEEEtran}
\bibliography{refs}

\end{document}